\documentclass[preprint,10pt,aps,notitlepage,onecolumn,a4paper]{revtex4}%
\usepackage{amssymb}
\usepackage{amsfonts}
\usepackage{amsmath}
\usepackage{graphicx}%
\begin{document}
\preprint{CERN-PH-TH/2010-123 and UMTG - 15}
\title[LMP]{Logistic Map Potentials}
\author{Thomas Curtright$^{\S }$}
\affiliation{Department of Physics, University of Miami, Coral Gables, FL 33124-8046, USA}
\author{Andrzej Veitia$^{\sharp}$}
\affiliation{Department of Physics, University of Miami, Coral Gables, FL 33124-8046, USA}
\pacs{PACS number}

\begin{abstract}
We develop and illustrate methods to compute all single particle potentials
that underlie the logistic map, $x\mapsto sx\left(  1-x\right)  $ for
$0<s\leq4$. \ We show that the switchback potentials can be obtained from the
primary potential through functional transformations. \ We are thereby able to
produce the various branches of the corresponding analytic potential
functions, which have an infinite number of branch points for generic $s>2$.
\ We illustrate the methods numerically for the cases $s=5/2$ and $s=10/3$.

\end{abstract}
\maketitle

\section{Introduction}

In two previous papers \cite{CZ}\ it was shown how functions defined on a
discrete lattice of time points may be smoothly interpolated in $t$, for a
continuum of time points, through the use of solutions to Schr\"{o}der's
nonlinear functional equation \cite{S}. \ If the effect of the first discrete
time step is given as the map
\begin{equation}
x\mapsto f_{1}\left(  x,s\right)  \ , \label{Map}%
\end{equation}
for some parameter $s$, then Schr\"{o}der's functional equation is%
\begin{equation}
s\Psi\left(  x,s\right)  =\Psi\left(  f_{1}\left(  x,s\right)  ,s\right)  \ ,
\label{SE}%
\end{equation}
with $\Psi$ to be determined. \ So, $f_{1}\left(  x,s\right)  =\Psi
^{-1}\left(  s\Psi\left(  x,s\right)  ,s\right)  $. \ A continuous
interpolation between the integer lattice of time points is then, for
\emph{any} $t$,%
\begin{equation}
f_{t}\left(  x,s\right)  =\Psi^{-1}\left(  s^{t}\Psi\left(  x,s\right)
,s\right)  \ . \label{Interpolation}%
\end{equation}
This can be a well-behaved, analytic and single-valued function of both $x$
\emph{and} $t$ provided that $\Psi^{-1}\left(  x,s\right)  $ is a
well-behaved, analytic, single-valued function of $x$,\ even though
$\Psi\left(  x,s\right)  $ might be, and typically is, multi-valued. \ In this
sense, analyticity in $x$ leads to analyticity in $t$.

As discussed in \cite{CZ}, the interpolation can be envisioned as the
trajectory of a particle passing through the initial $x$,
\begin{equation}
x\left(  t\right)  =f_{t}\left(  x,s\right)  \ , \label{Trajectory}%
\end{equation}
where the particle is moving according to Hamiltonian dynamics under the
influence of a potential, $V$.\ \ Up to additive and multiplicative constants,
at various times during the evolution of the particle, we have
\begin{equation}
V\left(  x\left(  t\right)  \right)  =-\left(  \frac{dx\left(  t\right)  }%
{dt}\right)  ^{2}\ . \label{Potential}%
\end{equation}
At $t=0$ this becomes\ $V\left(  x\right)  =-\left(  \ln^{2}s\right)  ~\left(
\frac{\Psi\left(  x,s\right)  }{d\Psi\left(  x,s\right)  /dx}\right)  ^{2}$.
\ Thus, $V$ may inherit multi-valuedness from $\Psi$ \cite{CZ}.

At other times the $x$ dependence of the potential also follows from that of
the velocity profile of the interpolation, $dx\left(  t\right)  /dt$, when the
latter is expressed as a function of $x\left(  t\right)  $. \ In general, this
will exhibit the branches of the underlying analytic potential function. \ But
more importantly for our purposes here, the various branches of the potential
can also be determined directly from the functional equation $V$ inherits from
$\Psi$. \ This functional equation is%
\begin{equation}
V\left(  f_{1}\left(  x,s\right)  ,s\right)  =\left(  \frac{d}{dx}f_{1}\left(
x,s\right)  \right)  ^{2}V\left(  x,s\right)  \ . \label{VFunctionalEqn}%
\end{equation}
If the map (\ref{Map}) possesses a fixed point, we may attempt to solve this
functional equation for $V$ by series in $x$ about that fixed point. \ We
shall discuss in some detail the circumstances for which this series method is
successful in the context of the logistic map $x\mapsto sx\left(  1-x\right)
$. \ In general, if the series can be constructed, it will of course have a
finite radius of convergence. \ However, the series result can then be
continued to other $x$ by making use of the functional equation itself (a
technique very familiar, e.g., for the $\Gamma$ and $\zeta$ functions) and
also by exploiting other special features for specific maps (cf.
$s\rightarrow2-s$ duality for the logistic map, discussed in Appendix A).
\ These additional techniques will allow us to construct convergent series
approximations for all branches of the potential in those situations where $V$
is multi-valued. \ The net result is a family of potential sequences that
encode for the corresponding continuous particle trajectories the various
fixed points, bifurcations, limit cycles, and chaotic behavior of the discrete
logistic map, for all $s$ of interest.

\vfill

\noindent\hrulefill

$^{\S }$curtright@miami.edu$\ \ \ \ \ ^{\sharp}$%
aveitia@physics.miami.edu\newpage

\section{Functional methods and series solutions}

Consider in detail the logistic map \cite{M,CE,F,K,C} on the unit interval,
$x\in\left[  0,1\right]  $,
\begin{equation}
x\mapsto sx\left(  1-x\right)  \ .
\end{equation}
For the most part, we restrict our considerations to parameter values
$s\in\left[  0,4\right]  $. \ The maximum of the map is $s/4$,\ obtained from
$x=1/2$, so without loss of any essential features, we will also most often
restrict $x\in\left[  0,s/4\right]  $. \ The map has fixed points at $x=0$ and
at $x_{\ast}=1-1/s$. \ Schr\"{o}der's equation for this map is \
\begin{equation}
s\Psi\left(  x,s\right)  =\Psi\left(  sx\left(  1-x\right)  ,s\right)  \ ,
\label{SFE}%
\end{equation}
and from this follows the functional equation for the underlying potential,
\begin{equation}
V\left(  sx\left(  1-x\right)  ,s\right)  =s^{2}\left(  1-2x\right)
^{2}V\left(  x,s\right)  \ . \label{VFE}%
\end{equation}
Applying the method of series solution about $x=0$, with initial conditions
that correspond to those used in \cite{CZ} for the function $\Psi=x+\frac
{1}{s-1}x^{2}+\cdots$, namely, $V\left(  0,s\right)  $ $=0,\ V^{\prime}\left(
0,s\right)  $ $=0,$ and $V^{\prime\prime}\left(  0,s\right)  =-2\ln^{2}s$, we
find:%
\begin{align}
V\left(  x,s\right)   &  =-\left(  \ln^{2}s\right)  ~U\left(  x,s\right)
\ ,\label{VtoU}\\
U\left(  x,s\right)   &  =x^{2}\left(  1+\sum_{n=1}^{\infty}a_{n}\left(
s\right)  ~x^{n}\right)  \ ,\ \ \ a_{1}=\frac{2}{1-s}\ ,\ \ \ a_{2}%
=\frac{5-3s}{\left(  s-1\right)  ^{2}\left(  s+1\right)  }\ ,\ \ \ \cdots\ .
\label{USeries}%
\end{align}
The higher coefficients in the expansion are determined recursively by%
\begin{equation}
a_{n+2}=\frac{1}{\left(  1-s^{n+2}\right)  }\left(  4a_{n+1}-4a_{n}%
+\sum_{j=1+\left\lfloor \frac{n-1}{2}\right\rfloor }^{n+1}\left(  -1\right)
^{n-j}a_{j}s^{j}\binom{j+2}{n+2-j}\right)  \text{ \ \ for }n\geq1\text{,}
\label{UCoefficientRecursion}%
\end{equation}
where $\left\lfloor \cdots\right\rfloor $ is the floor function. \ In
principle, this series solves (\ref{VFE})\ for any $s$, within its radius of
convergence. \ 

Based on numerical studies, we infer the radius of convergence \cite{Radius}
of the series depends on $s$ as follows: \
\begin{equation}
R\left(  s\right)  =\frac{1}{\lim\limits_{n\rightarrow\infty}\sup\left(
\left\vert a_{n}\left(  s\right)  \right\vert ^{1/n}\right)  }=\left\{
\begin{array}
[c]{ccc}%
\dfrac{1}{2} & \text{if} & 0<s\leq\frac{2}{3}\ ,\\
&  & \\
\left\vert 1-\dfrac{1}{s}\right\vert  & \text{if} & \frac{2}{3}\leq
s\leq2\ ,\\
&  & \\
\dfrac{s}{4} & \text{if} & 2\leq s\leq4\ .
\end{array}
\right.  \label{Radii}%
\end{equation}
For $0<s\leq2/3$, and also for $2\leq s\leq4$, the $\left\vert a_{n}%
\right\vert $ are monotonic for large $n$ and the radius immediately follows
either from the $\lim\sup$ expression in (\ref{Radii})\ or from the simple
ratio test, $R\left(  s\right)  =\lim\limits_{n\rightarrow\infty}~\left\vert
a_{n-1}\left(  s\right)  /a_{n}\left(  s\right)  \right\vert $. \ But for
$2/3<s<2$, there is spiky behavior in $\left\vert a_{n-1}/a_{n}\right\vert $
for intervals in $n$, which makes it difficult to use the simple ratio test to
determine $R$. \ This is because the $\left\vert a_{n}\left(  s\right)
\right\vert $ are not monotonic functions of $n$ for these values of $s$.
\ Occasionally the coefficients become small before changing sign.
\ Fortunately, the $\lim\sup$ expression for $R$ circumvents this spiky
behavior to yield the values given in (\ref{Radii}) for all $s$.

It is not difficult to work out explicit series results for $U\left(
x,s\right)  $, for generic $s$, say to $O\left(  x^{12}\right)  $. \ Such
polynomials in $x$ are sufficient approximations to obtain the graphics to
follow, when augmented with functional methods to be described. \ Based on
those explicit results, we infer that the series involve numerator
polynomials, $p_{n}\left(  s\right)  $, of order $1+\left(  n-2\right)
\left(  n-1\right)  /2$ in $s$, as well as \textquotedblleft
s-factorials\textquotedblright\ in the following form:%
\begin{equation}
\left(  s-1\right)  ^{2}U\left(  x,s\right)  =x^{2}\left(  \left(  s-1\right)
^{2}-2\left(  s-1\right)  x+\sum_{n=2}^{\infty}\frac{p_{n}\left(  s\right)
}{\left[  n\right]  _{s}!}~x^{n}\right)  \ .
\end{equation}
Here, deformed integers and factorials are defined by: $\ \left[  k\right]
_{s}=\frac{s^{k}-1}{s-1}$, and $\left[  n\right]  _{s}!=\prod_{k=1}^{n}\left[
k\right]  _{s}$. \ The recursion relation for the polynomials follows from
that for $a_{n}\left(  s\right)  $. \ It involves a mix of ordinary and
deformed integers:%
\begin{equation}
p_{n+2}\left(  s\right)  =\frac{1}{1-s}\left(  4p_{n+1}\left(  s\right)
-4\left[  n+1\right]  _{s}~p_{n}\left(  s\right)  +\left[  n+1\right]
_{s}!\sum_{j=1+\left\lfloor \frac{n-1}{2}\right\rfloor }^{n+1}\frac{\left(
-1\right)  ^{n-j}}{\left[  j\right]  _{s}!}~\binom{j+2}{n+2-j}~s^{j}%
~p_{j}\left(  s\right)  \right)  \ . \label{NumeratorPolyRecursion}%
\end{equation}
This recursion relation is seeded by%
\begin{equation}
p_{1}\left(  s\right)  =2\left(  1-s\right)  \ ,\ \ \ p_{2}\left(  s\right)
=5-3s\ .
\end{equation}
As written, it looks rather miraculous that the $\frac{1}{1-s}$ prefactor in
(\ref{NumeratorPolyRecursion}) is always canceled. \ Nevertheless, it is.
\ This follows from $\lim_{s\rightarrow1}p_{n}\left(  s\right)  $, but we have
not yet determined an elegant proof of this fact.

As originally obtained by Schr\"{o}der, there are three closed-form solutions
known for $\Psi$, for $s=-2,$ $2,$ and $4$. \ These are:%
\begin{align}
\Psi\left(  x,-2\right)   &  =\frac{\sqrt{3}}{6}\left(  2\pi-3\arccos\left(
x-\frac{1}{2}\right)  \right)  \ ,\ \ \ \Psi^{-1}\left(  x,-2\right)
=\frac{1}{2}-\cos\left(  \frac{2x}{\sqrt{3}}+\frac{\pi}{3}\right)
\ ,\nonumber\\
\Psi\left(  x,2\right)   &  =-\frac{1}{2}\ln\left(  1-2x\right)
\ ,\ \ \ \Psi^{-1}\left(  x,2\right)  =\frac{1}{2}\left(  1-e^{-2x}\right)
\ ,\nonumber\\
\Psi\left(  x,4\right)   &  =\left(  \arcsin\sqrt{x}\right)  ^{2}%
\ ,\ \ \ \Psi^{-1}\left(  x,4\right)  =\left(  \sin\sqrt{x}\right)  ^{2}\ .
\label{ExactCases}%
\end{align}
Note that while the $\Psi$ are multi-valued the inverse functions are all
single-valued. \ The corresponding closed-form expressions for
\begin{equation}
U\left(  x,s\right)  =\left(  \frac{\Psi\left(  x,s\right)  }{d\Psi\left(
x,s\right)  /dx}\right)  ^{2} \label{UQuadratic}%
\end{equation}
are also multi-valued and follow immediately:%
\begin{align}
U\left(  x,s=-2\right)   &  =\frac{1}{36}\left(  1+2x\right)  \left(
3-2x\right)  \left(  2\pi-3\arccos\left(  x-\frac{1}{2}\right)  \right)
^{2}\ ,\\
U\left(  x,s=2\right)   &  =\frac{1}{4}\left(  1-2x\right)  ^{2}\ln^{2}\left(
1-2x\right)  \ ,\label{U2}\\
U\left(  x,s=4\right)   &  =x\left(  1-x\right)  \arcsin^{2}\sqrt{x}\ .
\label{U4}%
\end{align}
Were we to start from these expressions for $U$, we could recover $\Psi$ by
solving and integrating (\ref{UQuadratic}). \ Another way to express the
result for $s=4$ is similar in form to that for $s=-2$, namely,%
\begin{equation}
U\left(  x,s=4\right)  =\frac{1}{4}x\left(  1-x\right)  \left(  \pi
-\arccos\left(  2x-1\right)  \right)  ^{2}\ .
\end{equation}
Indeed, it is well-known that the logistic maps for $s=4$ and $s=-2$ are
intimately related through the functional conjugacy of the underlying
Schr\"{o}der equations. \ (See Appendix A.)

\section{Obtaining the switchback potentials}

The sequence of switchback potentials, i.e. the various branches of the
analytic potential function, can be obtained from the functional equation for
the potential. \ (The same procedure also works to give the branch structure
of the Schr\"{o}der $\Psi$ function. \ For an example, see Appendix B.) \ This
follows from (\ref{SFE}) and (\ref{UQuadratic}), namely,%
\begin{equation}
U\left(  sx\left(  1-x\right)  \right)  =s^{2}\left(  1-2x\right)
^{2}U\left(  x\right)  \ . \label{UFunctionalEqn}%
\end{equation}
Next we write $y=sx\left(  1-x\right)  $, so then $x_{\pm}=\frac{1}{2}\left(
1\pm\sqrt{1-4y/s}\right)  $ and $\left(  1-2x_{\pm}\right)  ^{2}=1-4y/s$.
\ Thus the previous relation is just $U\left(  y\right)  =s\left(
s-4y\right)  U\left(  x_{\pm}\right)  $. \ Now, we rename $y\rightarrow x$ to
obtain%
\begin{equation}
x_{\pm}=\frac{1}{2}\pm\frac{1}{2}\sqrt{1-4x/s}\ ,
\end{equation}%
\begin{equation}
U_{\pm}\left(  x\right)  \equiv s\left(  s-4x\right)  ~U\left(  x_{\pm
}\right)  =s\left(  s-4x\right)  ~U\left(  \frac{1}{2}\pm\frac{1}{2}%
\sqrt{1-4x/s}\right)  \ .
\end{equation}
One of these potentials ($U_{-}$) reproduces the original series for the
primary potential expanded about $x=0$, as may be seen by direct comparison of
the series or by numerical evaluation. \ Alternatively, the other ($U_{+}$)
gives the potential on another sheet of the function's Riemann surface, hence
the first switchback potential for $s>2$. \ This is easily checked against the
closed-form results for the $s=4$ case. \ 

Since $U_{-}$ has built into it zeroes at both $x=0$ and $x_{\text{max}}=s/4$,
it is actually a more useful form than just the series about $x=0$. \ When
$2<s\leq4$, the series has radius of convergence $R\left(  s\right)  =s/4$.
\ However, from using $U_{-}$ instead of the direct series, we have
convergence over the whole closed interval, $x\in\left[  0,s/4\right]  $,
since then $0\leq\frac{1}{2}-\frac{1}{2}\sqrt{1-4x/s}\leq1/2<R\left(
s\right)  $ when $2<s\leq4$. \ Thus we need only evaluate $U$ appearing in
$U_{-}$ within the region of convergence of its series about zero. \ That is
to say, first construct the series for $U$, and then from that series build
$U_{-}$. \ Finally, identify this with $U_{0}$, the primary potential in the
sequence. \ 

Similarly, $U_{+}$ may be identified with the first switchback in the
sequence, $U_{1}$, but for better convergence properties it is useful to
build\ $U_{+}$\ from $U_{0}$ instead of $U$.\ \ By doing this when $2<s\leq4$,
we have convergence over the whole closed sub-interval, $x\in\left[  \frac
{1}{16}s^{2}\left(  4-s\right)  ,s/4\right]  $, with zeroes of $U_{1}%
$\ built-in at the end-points of the interval. \ 

So the primary potential and the first switchback are well-represented by%
\begin{align}
U_{0}\left(  x\right)   &  =s\left(  s-4x\right)  ~U\left(  \frac{1}{2}%
-\frac{1}{2}\sqrt{1-4x/s}\right)  \ ,\label{U0}\\
U_{1}\left(  x\right)   &  =s\left(  s-4x\right)  ~U_{0}\left(  \frac{1}%
{2}+\frac{1}{2}\sqrt{1-4x/s}\right)  \ . \label{U1}%
\end{align}
In the first of these expressions, $U$ is given by the direct series solution
of the functional equation (\ref{UFunctionalEqn}).

Now, go through this procedure all over again, beginning with $U\left(
sy\left(  1-y\right)  \right)  =s^{2}\left(  1-2y\right)  ^{2}U\left(
y\right)  $. \ Let $z=sy\left(  1-y\right)  $. \ Then $y_{\pm}=\frac{1}%
{2}\left(  1\pm\sqrt{1-4z/s}\right)  $ and $U\left(  z\right)  =s\left(
s-4z\right)  U\left(  y_{\pm}\right)  =s\left(  s-4z\right)  s\left(
s-4y_{\pm}\right)  U\left(  \left(  x_{\pm}\right)  _{\pm}\right)  $ where
$x_{\pm\pm}=\frac{1}{2}\left(  1\pm\sqrt{1-4y_{\pm}/s}\right)  $ with any
combination of $\pm$s allowed. $\ $Therefore $U\left(  z\right)  =s\left(
s-4z\right)  s\left(  s-2\left(  1\pm\sqrt{1-4z/s}\right)  \right)  $
$\times\ U\left(  \frac{1}{2}\pm\frac{1}{2}\sqrt{1-\frac{2}{s}\left(
1\pm\sqrt{1-4z/s}\right)  }\right)  $. \ Again, rename $z\rightarrow x$ to
obtain%
\begin{equation}
U_{\pm\pm}\left(  x\right)  =s\left(  s-4x\right)  U_{\pm}\left(  x_{\pm
}\right)  =s\left(  s-4x\right)  s\left(  s-2\left(  1\pm\sqrt{1-4x/s}\right)
\right)  U\left(  \frac{1}{2}\pm\frac{1}{2}\sqrt{1-\frac{2}{s}\left(
1\pm\sqrt{1-4x/s}\right)  }\right)  \ ,
\end{equation}
with any combination of $\pm$s allowed. \ And to improve the convergence
properties of this expression, replace $U$ on the RHS with $U_{0}$. \ 

This process may be continued indefinitely, through successive application of
the basic substitution $U\left(  x\right)  \rightarrow s\left(  s-4x\right)
~U\left(  x_{\pm}\right)  $. \ For example, the next set of potentials in the
sequence is $U_{\pm\pm\pm}\left(  x\right)  =s\left(  s-4x\right)  U_{\pm\pm
}\left(  x_{\pm}\right)  $, etc. \ In general, the $n$th iteration of the
procedure gives%
\begin{equation}
U_{\underset{\text{n times}}{\underbrace{\pm\pm\ldots\pm}}}\left(  x\right)
=s\left(  s-4x\right)  U_{\underset{\text{n-1 times}}{\underbrace{\pm\pm
\ldots\pm}}}\left(  x_{\pm}\right)  \ .
\end{equation}
Finally, at each iteration, we must select appropriate switchback potentials
out of the $2^{n}$ different expressions. \ In particular, we note that many
of the $U_{\pm\pm\ldots\pm}\left(  x\right)  $ will be complex-valued for the
$x$ intervals under consideration, and therefore they are not of immediate
interest since they do not govern the particle's evolution along the real
axis. (The continuation of the particle trajectory into the complex\ plane is
outside the scope of this paper.)

\section{Numerical examples}

\subsection{The potential sequence for $2<s\leq3$}

These are values of the parameter for which the discrete logistic map
converges to a single fixed point: \ There are no bifurcations.
\ Nevertheless, there are two sign choices when the functional equation for
the potential is applied once, four choices when it is applied twice, and so
on.%
\begin{align}
V_{\pm}\left(  x,s\right)   &  =s\left(  s-4x\right)  V_{0}\left(  \frac{1}%
{2}\pm\sqrt{\frac{1}{4}-\frac{x}{s}},s\right) \\
V_{\pm\pm}\left(  x,s\right)   &  =s\left(  s-4x\right)  s\left(  s-2\left(
1\pm\sqrt{1-4x/s}\right)  \right)  V_{0}\left(  \frac{1}{2}{\LARGE \pm}%
\sqrt{\frac{1}{4}-\frac{1}{s}\left(  \frac{1}{2}\pm\sqrt{\frac{1}{4}-\frac
{x}{s}}\right)  },s\right)  \ .
\end{align}
As it turns out from numerical studies, for $2<s\leq3$ only the positive sign
choices are needed to produce the sequence of switchback potentials.%
\begin{equation}
V_{1}\left(  x,s\right)  =s\left(  s-4x\right)  V_{0}\left(  \frac{1}{2}%
+\sqrt{\frac{1}{4}-\frac{x}{s}},s\right)  \ ,\ \ \ V_{2}\left(  x,s\right)
=s\left(  s-4x\right)  V_{1}\left(  \frac{1}{2}+\sqrt{\frac{1}{4}-\frac{x}{s}%
},s\right)  \ ,
\end{equation}
etc. \ In general, there is a recursion relation,
\begin{equation}
V_{n+1}\left(  x,s\right)  =s\left(  s-4x\right)  V_{n}\left(  \frac{1}%
{2}+\sqrt{\frac{1}{4}-\frac{x}{s}},s\right)  \ .
\label{FirstPotentialSequence}%
\end{equation}
The evolving particle moves through this sequence of potentials in succession,
with the potential index incrementing up by one each time the particle
encounters a turning point.

Consider the specific case $s=5/2$. \ This is representative for $2<s\leq3$.
\ For this case, the turning points converge onto the nontrivial fixed
point\ $x_{\ast}=1-1/s=3/5$. \ This is evident in the following graphs.%

\begin{center}
\includegraphics[
height=3.0095in,
width=4.5363in
]%
{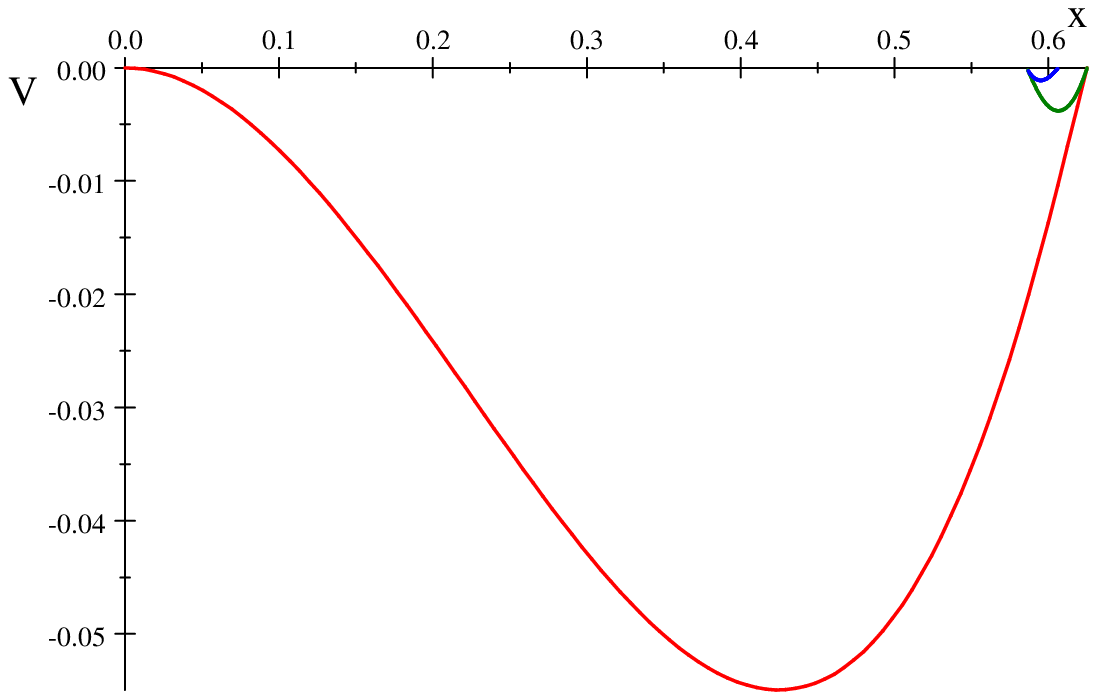}%
\\
$V_{0}$ in red, $V_{1}$ in green, and $V_{2}$ in blue, for $s=5/2$. \ The
first upper turning point is $x=5/8$. \ Subsequent lower and upper turning
points are obtained just by iterating the $s=5/2$ logistic map, starting with
$x=5/8$, and are shown on the next magnified graph as colored points.
\end{center}
\begin{center}
\includegraphics[
height=3.0095in,
width=4.5363in
]%
{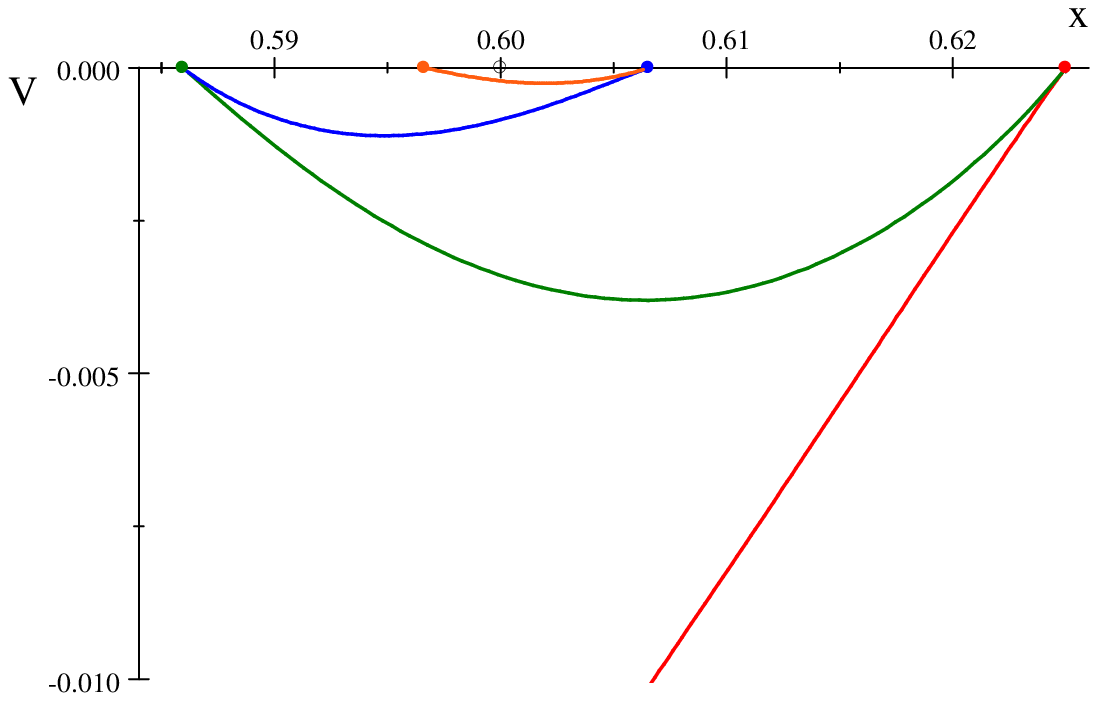}%
\\
$V_{0}$ in red, $V_{1}$ in green, $V_{2}$ in blue, and $V_{3}$ in orange, for
$s=5/2$. \ The first upper turning point is $x=5/8$, the first lower turning
point is $x=\frac{75}{128}=0.58594$, the second upper turning point is
$x=\frac{19\,875}{32\,768}=0.60654$, etc., as obtained by map iteration. \ The
nontrivial fixed point is at $x_{\ast}=3/5$, indicated by the black circle on
the $x$ axis.
\end{center}
As the zero-energy particle moves through this sequence of increasingly
shallow, narrowing potentials, its average speed decreases, giving the
\emph{appearance} of a dissipative system. \ Nevertheless, even as the
particle motion subsides upon convergence into the fixed point at $x=3/5$,
energy is rigorously conserved through changes in the potential.

Insofar as the turning points are branch points for the corresponding analytic
potential function, and the various switchback potentials are just the values
of that analytic function on the various sheets of its Riemann surface, this
is convincing numerical evidence for that function to have an infinite number
of such branch points, for generic $s$. \ While of course there are many
well-known functions with this property (for example, the inverses of Bessel
functions, $J_{n}^{-1}\left(  x\right)  $), this would seem to account for the
historical fact that closed-form solutions have \emph{not} been found for
generic $s$, or even for specific $s$, except in those very special cases
where the number of branch points is one or two. Recall the exact closed-form
cases $s=2$ and $s=4$ have analytic potential functions with one and two
branch points, respectively, as evident in (\ref{U2}) and (\ref{U4}).

We may check numerically the transit times for each potential in the sequence:
$\ \Delta t_{n}=1$, for $n\geq1$, within the expected uncertainties for
truncation of the series (\ref{USeries}). \ For example, using Mathematica to
compute the initial series (\ref{USeries}) to 200th order and also to evaluate
the numerical integrations, we find$\ \int_{1/2}^{5/8}\frac{dx}{\sqrt{-V_{0}}%
}=1.000000$, $\int_{75/128}^{5/8}\frac{dx}{\sqrt{-V_{1}}}=1.000000$,
$\int_{75/128}^{19\,875/32\,768}\frac{dx}{\sqrt{-V_{2}}}=1.000000$,
$\int_{1281241875/2147483648}^{19\,875/32\,768}\frac{dx}{\sqrt{-V_{3}}%
}=1.000000$, etc., corresponding to the iterations\ $\left.  \frac{5}%
{2}x\left(  1-x\right)  \right\vert _{x=1/2}=\frac{5}{8}$, $\left.  \frac
{5}{2}x\left(  1-x\right)  \right\vert _{x=5/8}=\frac{75}{128}$, $\left.
\frac{5}{2}x\left(  1-x\right)  \right\vert _{x=75/128}=\frac{19\,875}%
{32\,768}$, $\left.  \frac{5}{2}x\left(  1-x\right)  \right\vert
_{x=19\,875/32\,768}=\frac{1281\,241\,875}{2147\,483\,648}$, etc. \ These
transit times are consistent with the continuously evolving, zero-energy
particle moving under the influence of each of the potentials in
succession,\ $V_{0}\rightarrow V_{1}\rightarrow V_{2}\rightarrow\cdots$,
thereby converging onto the fixed point at $x=3/5$. \ The numbers confirm that
for cases with $2<s\leq3$, only positive roots are needed to obtain all the
switchback potentials, as previously indicated in
(\ref{FirstPotentialSequence}).

\subsection{The potential sequence family for $3<s<4$}

These are values of the parameter for which the discrete logistic map produces
bifurcations and asymptotes to limit cycles, rather than unique fixed points,
and as $s$ is increased, chaotic behavior erupts. \ Generally speaking, for
these values of $s$ the analytic potentials driving the trajectories are more
complicated than for $s\leq3$, and their branches are not so easy to enumerate
as they are encountered by the evolving particle. \ The potentials here may be
grouped into a \emph{family of sequences}, where potentials with common upper
and lower turning points constitute a single member of the family. \ From one
family member to the next, for fixed sequence index, the potentials exhibit a
progressive shallowing (similar to the behavior of the individual potentials
for $s=5/2$). \ However, the $V$s for a given family member successively
deepen as the sequence index increases. \ Consequently, the locally averaged
speed of the particle moving through the potentials does not necessarily
diminish, but often increases, giving the \emph{appearance} of a driven
system. \ Nevertheless, energy is still rigorously conserved.

The limit cycle situation may be illustrated by the case $s=10/3$, while the
chaotic situation is aptly illustrated by the uppermost value, $s=4$. \ The
latter case was fully discussed in \cite{CZ}, and originally led to the notion
of the switchback potentials. \ On the other hand, the potential structure for
the $s=10/3$ case is more complicated than for $s=4$, the latter having only
one member for its family of potential sequences, and in fact $s=10/3$ is much
more representative of the situation for generic parameter values in the range
$3<s<4$. \ Therefore, we discuss $s=10/3$ in some detail. \ This case evolves
towards a two-cycle, consisting of the points $\frac{1}{20}\sqrt{13}+\frac
{13}{20}=\allowbreak0.830\,278\,$and $\frac{13}{20}-\frac{1}{20}\sqrt
{13}=\allowbreak0.469\,722\,$. \ The first few terms in the potential sequence
(\ref{FirstPotentialSequence}) are shown here, for this particular $s$.%
\begin{center}
\includegraphics[
height=3.0095in,
width=4.5363in
]%
{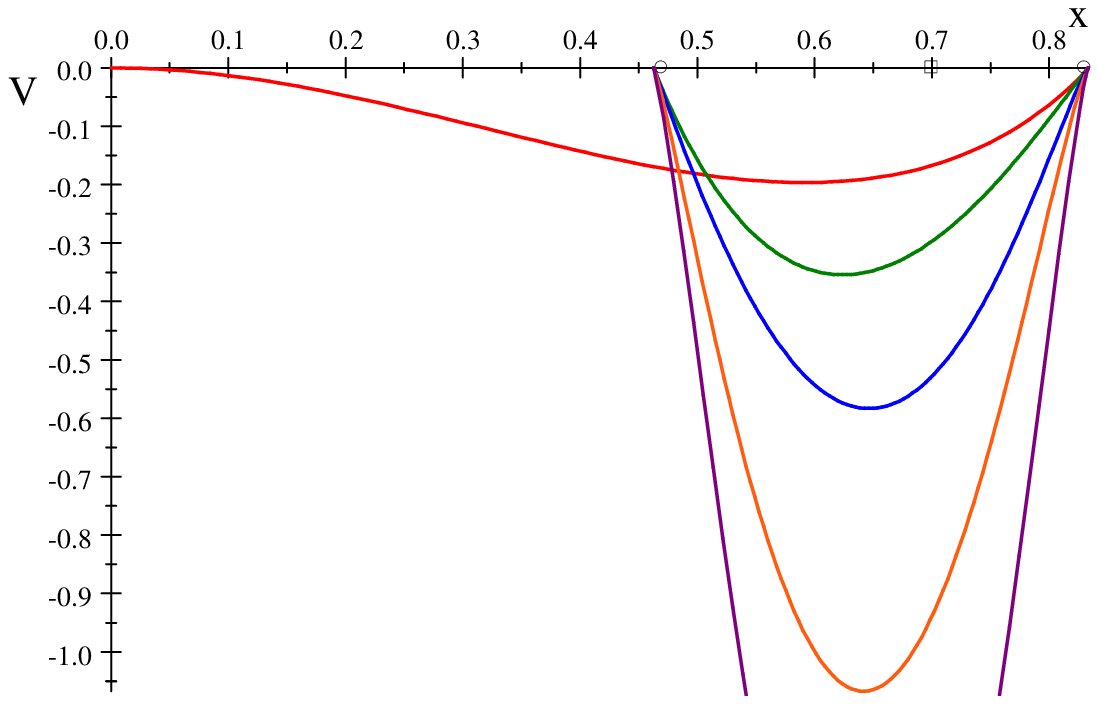}%
\\
$V_{0}$ in red, $V_{1}$ in green, $V_{2}$ in blue, $V_{3}$ in orange, and
$V_{4}$ in purple, for $s=10/3$. \ The upper turning point is $x=5/6$. \ The
two-cycle points are small black circles on the axis. \ The small black square
is the nontrivial fixed point of the map, $x_{\ast}=7/10$, and is revisited
after integer time steps, but it is not a fixed point for the continuous
evolution.
\end{center}
Deepening of the $V$s is evident for this \textquotedblleft mother
sequence.\textquotedblright\ \ The turning points for this sequence do
\emph{not} converge in this $s=10/3$ case. \ Rather, they are fixed at
$\frac{5}{6}$ for the upper turning point, and $\frac{25}{54}$ for the lower
turning point, the latter obtained by acting on the former with the
$s=\frac{10}{3}$ discrete logistic map : \ $\frac{5}{6}=\allowbreak
0.833\,333\,\mapsto\frac{25}{54}=\allowbreak0.462\,963\,$. \ While these
points are not far from the two-cycle points, as evident below, close does not
count (except when playing p\'{e}tanque).

\begin{center}%
{\parbox[b]{3.3798in}{\begin{center}
\includegraphics[
height=2.2432in,
width=3.3798in
]%
{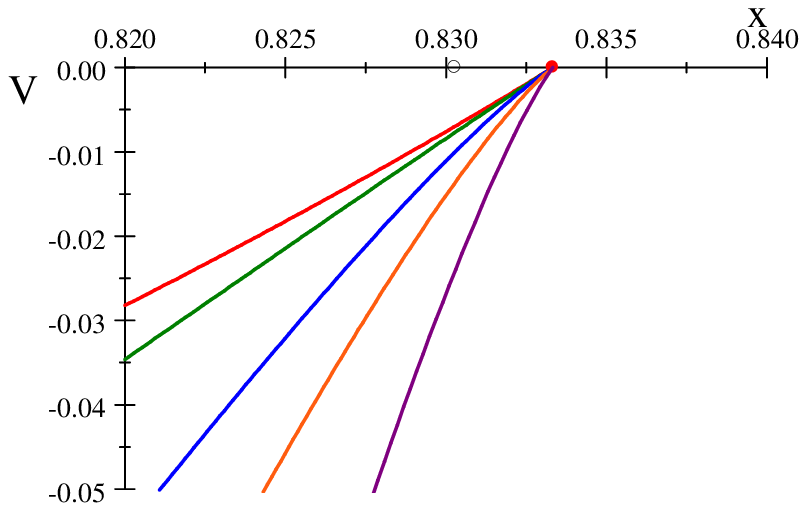}%
\\
$V_{0}$ red, $V_{1}$ green, $V_{2}$ blue, $V_{3}$ orange, \& $V_{4}$ purple,
for $s=10/3$. \ The upper turning point is $x=\frac{5}{6}$.
\end{center}}}%
{\parbox[b]{3.3798in}{\begin{center}
\includegraphics[
height=2.2432in,
width=3.3798in
]%
{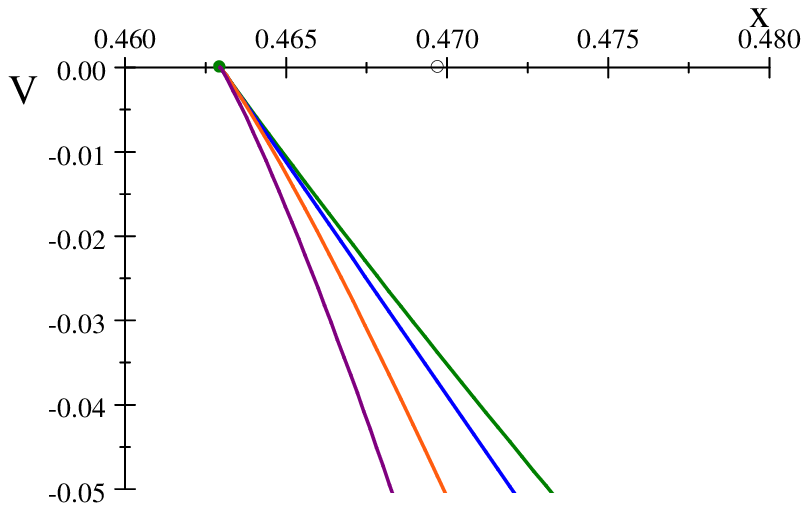}%
\\
$V_{1}$ green, $V_{2}$ blue, $V_{3}$ orange, \& $V_{4}$ purple, for $s=10/3$.
\ The lower turning point is $x=\frac{25}{54}$.
\end{center}}}%

\end{center}

\noindent Other sequences in the family of potentials are needed to achieve
convergence of the continuous particle trajectory, for unit time steps, onto
the two-cycle of the map.

So, we define a second sequence in the family of potentials by changing the
sign of the root in the arguments of the potentials in the first sequence,
while noting that $W_{0}\left(  x,s\right)  =s\left(  s-4x\right)
V_{0}\left(  \frac{1}{2}-\sqrt{\frac{1}{4}-\frac{x}{s}},s\right)  $ just
reproduces $V_{0}$. \ More interesting cases of these \textquotedblleft
offspring\textquotedblright\ potentials are given by
\begin{equation}
W_{n}\left(  x,s\right)  =s\left(  s-4x\right)  V_{n}\left(  \frac{1}{2}%
-\sqrt{\frac{1}{4}-\frac{x}{s}},s\right)  \ \text{, \ \ for \ \ }%
n\geq1\text{.}%
\end{equation}
These do \emph{not} reproduce any of the previous potentials. \ We plot a few
for the $s=10/3$ case.%
\begin{center}
\includegraphics[
height=3.0095in,
width=4.5363in
]%
{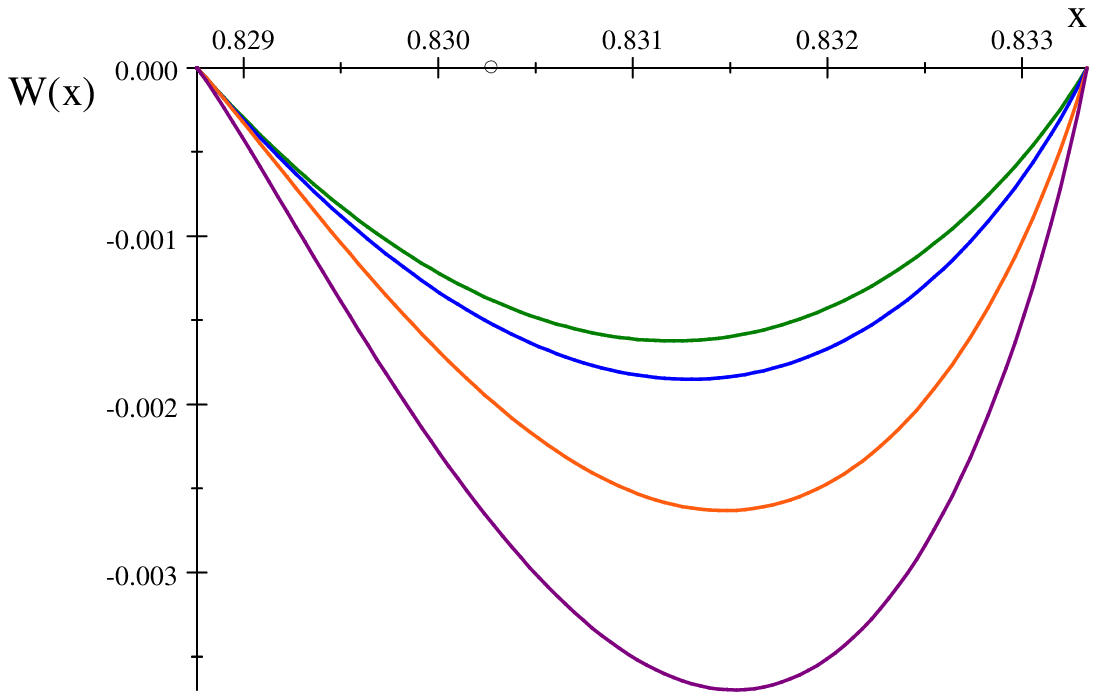}%
\\
$W_{1}$ green, $W_{2}$ blue, $W_{3}$ orange, and $W_{4}$ purple, for $s=10/3$.
\end{center}
For this second sequence the lower turning point can be obtained by acting
with the discrete map on the lower turning point for the first sequence, i.e.
$\frac{25}{54}=0.462\,963\mapsto\frac{3625}{4374}=\allowbreak0.828\,761$,
while the upper turning point remains the same as for the first sequence, i.e.
$\frac{5}{6}=\allowbreak0.833\,333$. \ Continuing the enumeration of the
potential family members, we define a third sequence of potentials, and we
plot a few.%
\begin{equation}
X_{n}\left(  x,s\right)  =s\left(  s-4x\right)  W_{n}\left(  \frac{1}{2}%
+\sqrt{\frac{1}{4}-\frac{x}{s}},s\right)  \ \text{, \ \ for \ \ }%
n\geq1\text{,}%
\end{equation}%
\begin{center}
\includegraphics[
height=3.0095in,
width=4.5363in
]%
{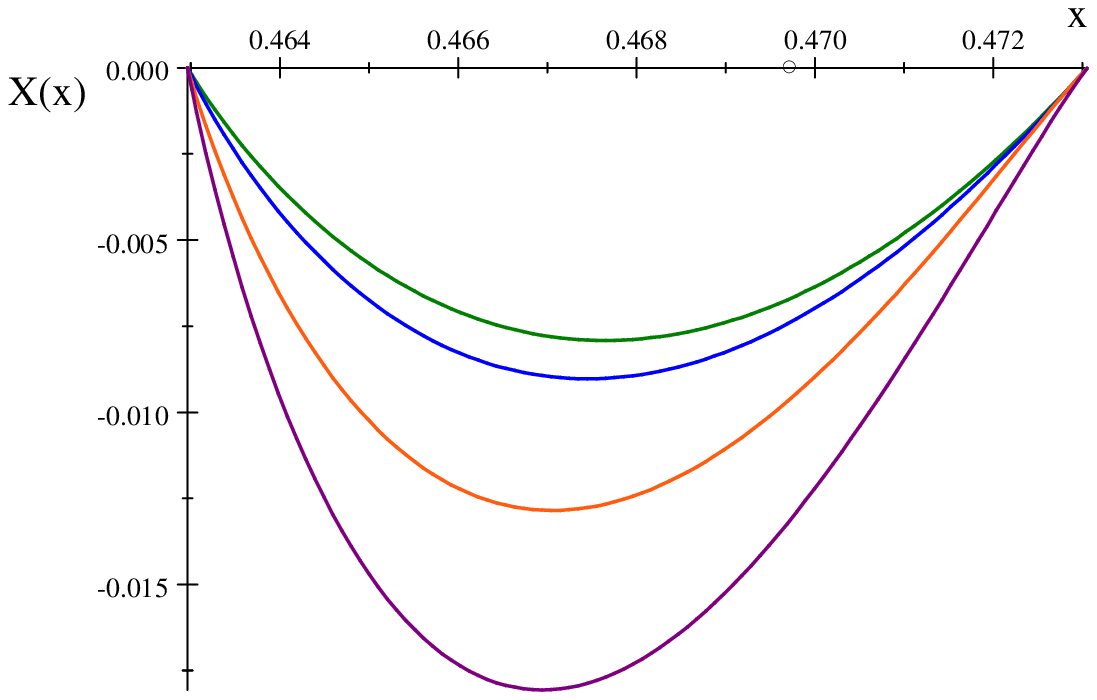}%
\\
$X_{1}$ green, $X_{2}$ blue, $X_{3}$ orange, and $X_{4}$ purple, for $s=10/3$.
\end{center}
For this third sequence the lower turning point remains the same as for the
first sequence, i.e. $\frac{25}{54}=0.462\,963$, while the upper turning point
can be obtained by acting with the discrete map on the lower turning point for
the second sequence, i.e. $\frac{3625}{4374}=\allowbreak0.828\,761\mapsto
\frac{13\,575\,625}{28\,697\,814}=\allowbreak0.473\,054$. \ 

This procedure may be continued indefinitely. \ We define fourth and fifth
sequences as$^{\dagger}$%
\begin{align}
Y_{n}\left(  x,s\right)   &  =s\left(  s-4x\right)  X_{n}\left(  \frac{1}%
{2}-\sqrt{\frac{1}{4}-\frac{x}{s}},s\right)  \ ,\\
Z_{n}\left(  x,s\right)   &  =s\left(  s-4x\right)  Y_{n}\left(  \frac{1}%
{2}+\sqrt{\frac{1}{4}-\frac{x}{s}},s\right)  \ ,
\end{align}
etc., and we plot the first few of these new sequences.%
\begin{center}
\includegraphics[
height=3.0095in,
width=4.5363in
]%
{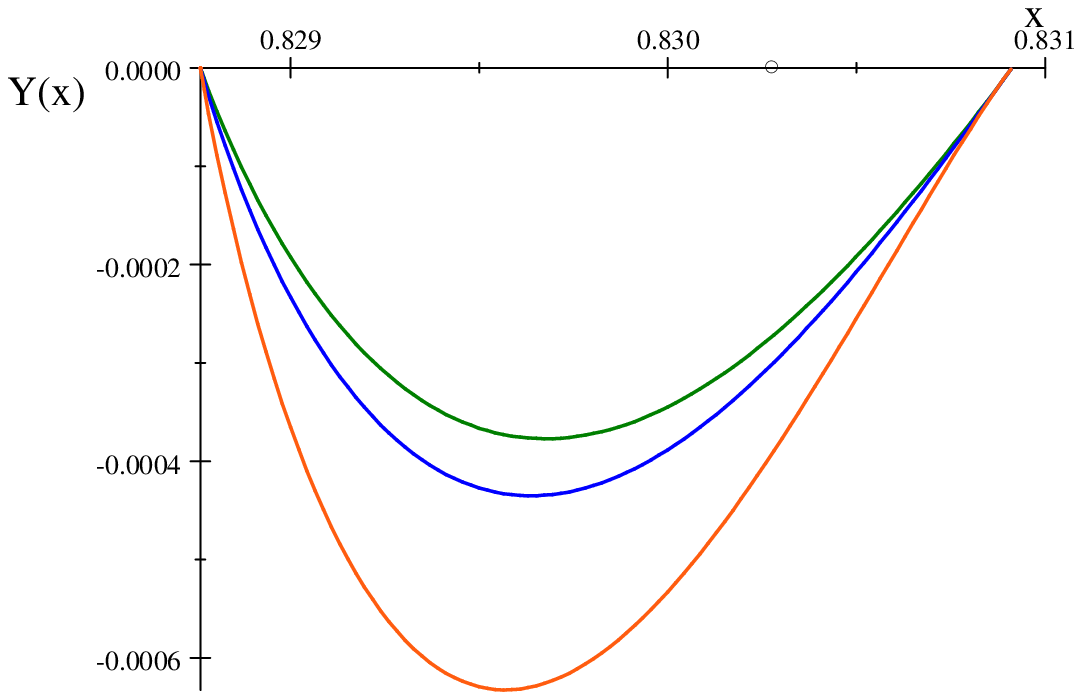}%
\\
$Y_{1}$ green, $Y_{2}$ blue, and $Y_{3}$ orange, for $s=10/3$.
\end{center}
\vfill

\noindent\hrulefill

$^{\dagger}$A more systematic notation would employ two potential indices, to
denote an infinite matrix of potentials. \ For example, we could designate
$V_{n}=V_{n}^{\left(  0\right)  }$, $W_{n}=V_{n}^{\left(  1\right)  }$,
$X_{n}=V_{n}^{\left(  2\right)  }$, $Y_{n}=V_{n}^{\left(  3\right)  }%
$,\ $Z_{n}=V_{n}^{\left(  4\right)  }$, etc. \ The super-index here is the
\emph{family number}.%
\begin{center}
\includegraphics[
height=3.0095in,
width=4.5363in
]%
{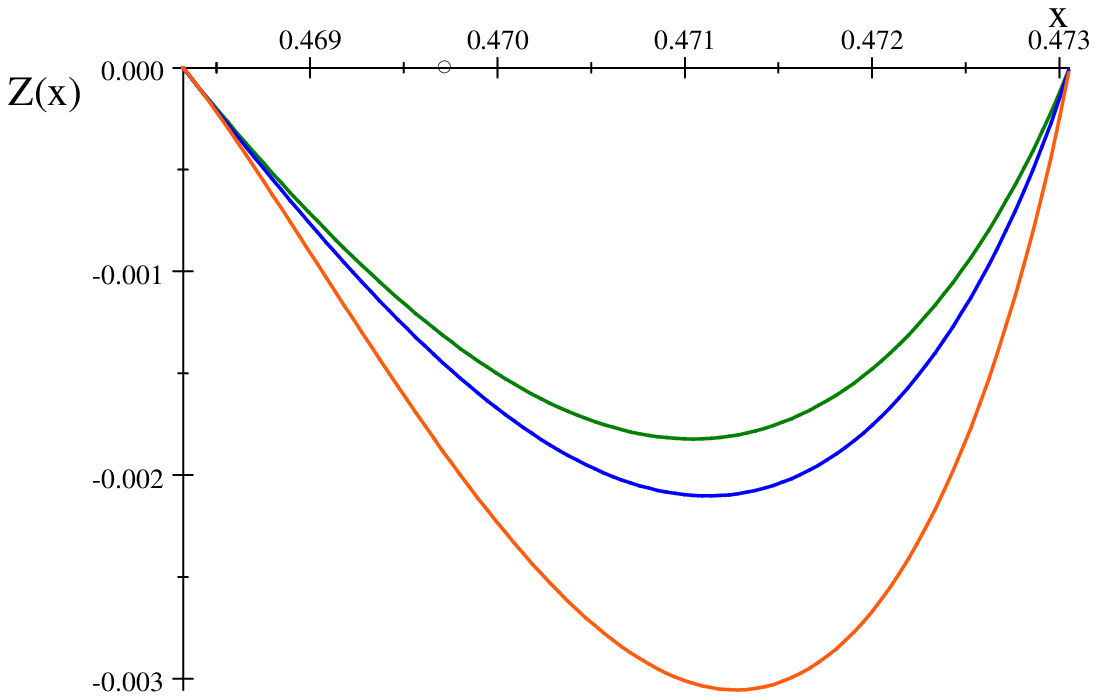}%
\\
$Z_{1}$ green, $Z_{2}$ blue, and $Z_{3}$ orange, for $s=10/3$.
\end{center}
The story here is somewhat analogous to, but more complicated than, the
situation for $2<s\leq3$. \ Instead of individual potentials in the sequence
$\left\{  V_{n}\right\}  $ converging onto the fixed point, as would be the
case for $2<s\leq3$, here each of the individual $V_{n}$ has proliferated into
a \textquotedblleft sideways\textquotedblright\ sequence of potentials,
$\left\{  V_{n},W_{n},X_{n},Y_{n},Z_{n},\cdots\right\}  $, which converge to
the points in the 2-cycle. \ To show what we mean by this, we plot $\left\{
V_{1},W_{1},Y_{1}\right\}  $ and $\left\{  V_{1},X_{1},Z_{1}\right\}  $ near
the lower and upper points, respectively, in the two-cycle.

\begin{center}%
{\parbox[b]{3.3798in}{\begin{center}
\includegraphics[
height=2.2582in,
width=3.3798in
]%
{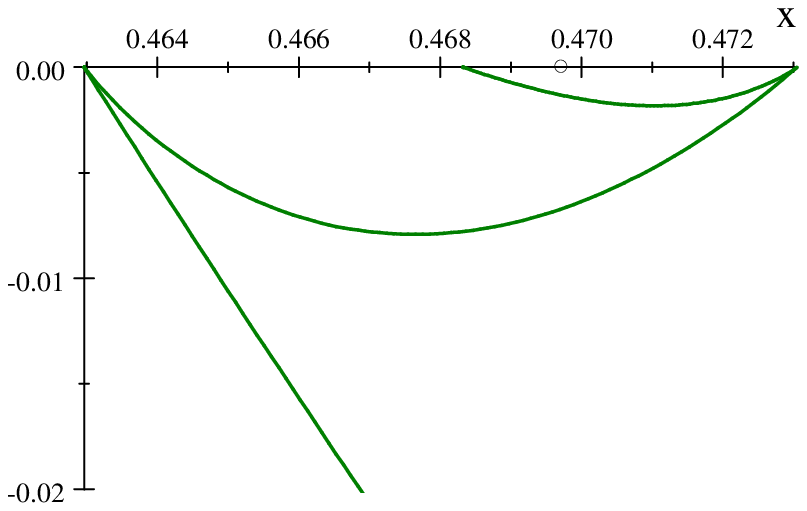}%
\\
$V_{1}$, $W_{1}$, and $Y_{1}$ near the lower point in the two-cycle for
$s=10/3$.
\end{center}}}%
{\parbox[b]{3.3798in}{\begin{center}
\includegraphics[
height=2.2582in,
width=3.3798in
]%
{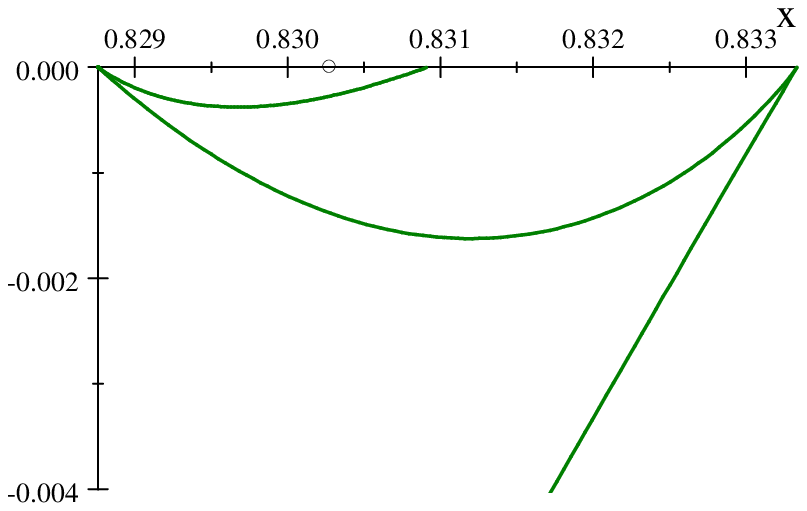}%
\\
$V_{1}$, $X_{1}$, and $Z_{1}$ near the upper point in the two-cycle for
$s=10/3$.
\end{center}}}%

\end{center}

\noindent This type of behavior is repeated by $\left\{  V_{n},W_{n}%
,Y_{n}\right\}  $ and $\left\{  V_{n},X_{n},Z_{n}\right\}  $ for each $n$.

To complete the picture for the evolving particle, it is necessary to
understand the order in which the potentials act, i.e. to determine how the
switches are made from one potential to another as the turning points are
encountered by the particle. \ Unlike the situation with $s\leq3$, here this
can be a bit tedious. \ From a numerical examination of various transit times
(and also from the branch structure of the underlying Schr\"{o}der auxiliary
function, as exhibited in Appendix B) we infer the order of the potentials for
the $s=10/3$ case to produce a \emph{chemin des \'{e}nergies potentielles}:%
\begin{equation}
V_{0}\rightarrow\underset{\Delta t=1}{\underbrace{V_{1}}}\rightarrow
\underset{\Delta t=1}{\underbrace{V_{2}\rightarrow W_{1}}}\rightarrow
\underset{\Delta t=1}{\underbrace{W_{2}\rightarrow V_{3}\rightarrow X_{1}}%
}\rightarrow\underset{\Delta t=1}{\underbrace{X_{2}\rightarrow V_{4}%
\rightarrow W_{3}\rightarrow Y_{1}}}\rightarrow\underset{\Delta t=1}%
{\underbrace{Y_{2}\rightarrow W_{4}\rightarrow V_{5}\rightarrow X_{3}%
\rightarrow Z_{1}}}\rightarrow\cdots\ , \label{chemin}%
\end{equation}
etc. \ Here we have also indicated how the potentials combine into groups with
unit total transit time. \ For example, again using Mathematica to compute the
initial series (\ref{USeries}) to 200th order and computing the transit times
numerically, we find: \ $\int_{25/54}^{5/6}\frac{dx}{\sqrt{-V_{1}}}=1.000000$;
$\int_{25/54}^{5/6}\frac{dx}{\sqrt{-V_{2}}}+\int_{3625/4374}^{5/6}\frac
{dx}{\sqrt{-W_{1}}}=0.825728+0.174272=1.000000$; $\int_{3625/4374}^{5/6}%
\frac{dx}{\sqrt{-W_{2}}}+\int_{25/54}^{5/6}\frac{dx}{\sqrt{-V_{3}}}%
+\int_{25/54}^{13575625/28697814}\frac{dx}{\sqrt{-X_{1}}}%
=0.164433+0.661295+0.174272=1.000000$; etc.

It is not difficult to extend (\ref{chemin}) by defining additional sequence
family members. \ Note the turning points must match-up for adjacent
potentials along the path. \ Also note the family number plus the sequence
number (i.e. $N=m+n$ for the potential matrix element $V_{n}^{\left(
m\right)  }$ of the last footnote) is the same for each potential belonging to
a group with total $\Delta t=1$, and this $N$ increments by one as the
particle moves from one potential group to the next. \ 

\section{Conclusion}

The point of view supported in this paper is that the logistic map, and other
discrete time-stepped dynamical models, may be regarded as continuously
evolving Hamiltonian systems sampled at integer times. \ For this view to be
valid, the continuous system must be allowed to undergo a series of
switchbacks whereupon the potential affecting the dynamics changes when the
evolving particle encounters a turning point. \ From a perspective of
configuration space covering manifolds, in the case of simple one-dimensional
motion, the particle moves from one sheet of a Riemann surface to another, to
experience a different branch of the underlying analytic potential.

The methods of this paper may be used directly to determine such branches of
the potential for the logistic map, for any value of $s$, as well as other
one-dimensional maps. \ A more extensive study of other examples is underway
\cite{CV}. \ While peculiar behavior is possible for exceptional maps (say,
for special values$^{\ddagger}$ of $s$), so far as we are aware, any such
behavior can always be analyzed using the potential framework presented here.

In total, for all parameter values governing a particular map, the collection
of potential sequence families constitute what we may call a \textquotedblleft
potential fractal\textquotedblright\ with self-similarities qualitatively
evident in the various graphs, as visible in the above. \ Perhaps such
potential fractals have a significant role to play in continuum physics.
\ Applications might involve any of the usual systems exhibiting chaotic
behavior \cite{C}, including accelerator beams \cite{Lee}, or perhaps
cosmological models \cite{CosmologyPotentials}.

\begin{acknowledgments}
\textit{We thank David Fairlie, Xiang Jin, Luca Mezincescu, and especially
Cosmas Zachos, for sharing their thoughts about functional evolution methods.
\ One of us (TC) thanks the CERN Theoretical Physics Group for its gracious
hospitality and generous support. \ The numerical calculations and graphics in
this paper were made using Maple}$^{\textregistered}$\textit{, Mathematica}%
$^{\textregistered}$, \textit{and MuPAD}$^{\textregistered}$\textit{. \ This
work was also supported by NSF Award 0855386.}
\end{acknowledgments}

\vfill

\noindent\hrulefill

$^{\ddagger}$A peculiar example is provided by $s=1$. \ For this case the
appropriate limit of (\ref{USeries}) is not convergent but rather an
asymptotic series. \ (See Appendix C.)\newpage

\section{Appendix A: \ Functional conjugation}

Functional conjugacy for the logistic map may be expressed in terms of a
linear function. \ Namely,%
\begin{equation}
g\left(  sx\left(  1-x\right)  \right)  =\left(  2-s\right)  g\left(
x\right)  \left(  1-g\left(  x\right)  \right)  \ ,
\end{equation}
where%
\begin{equation}
g\left(  z\right)  =\frac{1}{2-s}\left(  1-s+sz\right)  \ ,\ \ \ g^{-1}\left(
z\right)  =\frac{1}{s}\left(  s-1+\left(  2-s\right)  z\right)  \ .
\end{equation}
It is often useful to refer to these maps as $g\left(  z\right)  =g_{s}\left(
z\right)  $ and $g^{-1}\left(  z\right)  =g_{2-s}\left(  z\right)  $.
\ Pursuing the conjugacy a bit farther, let $f_{s}\left(  x\right)  =sx\left(
1-x\right)  $, then the conjugacy equation for the map can be written in
various ways:%
\begin{gather}
g\circ f_{s}=f_{2-s}\circ g\ ,\ \ \ f_{s}\circ g^{-1}=g^{-1}\circ f_{2-s}\ ,\\
g\circ f_{s}\circ g^{-1}=f_{2-s}\ ,\ \ \ g^{-1}\circ f_{2-s}\circ g=f_{s}\ .
\end{gather}
Moreover, Schr\"{o}der's equation is the functional composition%
\begin{equation}
s\circ\Psi=\Psi\circ f_{s}\ ,
\end{equation}
where $s:x\rightarrow sx$ is the simple multiplicative map. \ Under functional
conjugacy by $g$, the RHS of Schr\"{o}der's equation becomes
\begin{align}
g\circ\Psi\circ f_{s}\circ g^{-1}  &  =g\circ\Psi\circ g^{-1}\circ
f_{2-s}=\Psi_{g}\circ f_{2-s}\label{gConjugation1}\\
\Psi_{g}  &  \equiv g\circ\Psi\circ g^{-1} \label{gConjugation2}%
\end{align}

\subsection{Alternate fixed point expansions}

Consider again the logistic map, $x\mapsto sx\left(  1-x\right)  $. \ On the
one hand, with a subscript to distinguish other constructions to follow,
expansions about the trivial fixed point at $x=0$ stem from:%
\begin{align}
s\Psi_{0}\left(  x,s\right)   &  =\Psi_{0}\left(  sx\left(  1-x\right)
,s\right)  \ ,\\
\Phi_{0}\left(  sx,s\right)   &  =s\Phi_{0}\left(  x,s\right)  \left(
1-\Phi_{0}\left(  x,s\right)  \right)  \ .
\end{align}
For the inverse Schr\"{o}der functions, we consistently use the notation
$\Phi=\Psi^{-1}$. \ On the other hand, expansions about the non-trivial fixed
point at $x=1-1/s$ stem from:%
\begin{align}
\lambda\Psi_{\ast}\left(  z\right)   &  =\Psi_{\ast}\left(  \lambda z+\left(
\lambda-2\right)  z^{2}\right)  \ ,\\
\Phi_{\ast}\left(  \lambda z\right)   &  =\lambda\Phi_{\ast}\left(  z\right)
+\left(  \lambda-2\right)  \Phi_{\ast}^{2}\left(  z\right)  \ ,
\end{align}
where%
\begin{equation}
\lambda=2-s\ ,\ \ \ s=2-\lambda\ ,\ \ \ x=1-\frac{1}{s}+z\ ,\ \ \ z=x+\frac
{1-s}{s}\ .
\end{equation}
The two expansions produce \emph{the same functions}, only slightly disguised.
\ Here is the detailed relation between them, to be proved in the next
subsection.%
\begin{equation}
\Psi_{\ast}\left(  z\right)  =\frac{\lambda}{2-\lambda}\Psi_{0}\left(
\frac{2-\lambda}{\lambda}z,\lambda\right)  \ ,\text{ \ \ i.e. \ \ }\Psi_{\ast
}\left(  x+\frac{1-s}{s}\right)  =\frac{2-s}{s}\Psi_{0}\left(  \frac{s}%
{2-s}\left(  x+\frac{1-s}{s}\right)  ,2-s\right)  \ .
\end{equation}%
\begin{equation}
\Phi_{\ast}\left(  z\right)  =\frac{\lambda}{2-\lambda}\Phi_{0}\left(
\frac{2-\lambda}{\lambda}z,\lambda\right)  \ ,\text{ \ \ i.e. \ \ }\Phi_{\ast
}\left(  x+\frac{1-s}{s}\right)  =\frac{2-s}{s}\Phi_{0}\left(  \frac{s}%
{2-s}\left(  x+\frac{1-s}{s}\right)  ,2-s\right)  \ .
\end{equation}
These relations are an extension of the previously known functional conjugacy
that relates $s=4$ and $s=-2$. \ 

It is better notation to call the alternate series solutions $\Psi_{\ast
}\left(  x,s\right)  $ and $\Phi_{\ast}\left(  x,s\right)  $, instead of
$\Psi_{\ast}\left(  z\right)  $ and $\Phi_{\ast}\left(  z\right)  $. \ Then
the \textquotedblleft dual\textquotedblright\ parameter is
\begin{equation}
s_{\ast}=2-s\ , \label{sDual}%
\end{equation}
and the previous relations are more succinctly written as:%
\begin{equation}
\Psi_{\ast}\left(  x,s\right)  =\frac{s_{\ast}}{s}~\Psi_{0}\left(  \frac
{sx}{s_{\ast}},s_{\ast}\right)  \ ,\ \ \ \Phi_{\ast}\left(  x,s\right)
=\frac{s_{\ast}}{s}~\Phi_{0}\left(  \frac{sx}{s_{\ast}},s_{\ast}\right)  \ .
\end{equation}
That is to say, the series solution of Schr\"{o}der's equation for map
parameter $2-s$, about the trivial fixed point at $x=0$, is conjugate to the
series solution of Schr\"{o}der's equation for map parameter $s$, about the
nontrivial fixed point at $x_{\ast}=1-1/s$. \ Now, as previously noted,
functional conjugation by $g$ gives (\ref{gConjugation1}) and
(\ref{gConjugation2}). \ But then, $g\circ s\circ g^{-1}\left(  u\right)
=g\left(  s-1+\left(  2-s\right)  u\right)  =\frac{\left(  1-s\right)  ^{2}%
}{2-s}+su\ $, so
\begin{equation}
g\circ s\circ g^{-1}\circ\Psi_{g}\left(  w,s\right)  =\frac{\left(
1-s\right)  ^{2}}{2-s}+s\Psi_{g}\left(  w,s\right)  \ .
\end{equation}
That is to say, the $g$-conjugated Schr\"{o}der function obeys the
inhomogeneous functional equation,%
\begin{equation}
\frac{\left(  1-s\right)  ^{2}}{2-s}+s\Psi_{g}\left(  x,s\right)  =\Psi
_{g}\left(  \left(  2-s\right)  x\left(  1-x\right)  ,s\right)  \ ,
\end{equation}
whereas $\Psi_{\ast}$ obeys a homogeneous equation, but with shifted
position-dependent arguments.%
\begin{equation}
\left(  2-s\right)  \Psi_{\ast}\left(  x+\frac{1-s}{s},s\right)  =\Psi_{\ast
}\left(  sx\left(  1-x\right)  +\frac{1-s}{s},s\right)  \ .
\end{equation}

To summarize the relations between the various functions:
\begin{subequations}
\begin{align}
\Psi_{0}\left(  x,s\right)   &  =\frac{2-s}{s}~\Psi_{\ast}\left(  \frac
{sx}{2-s},2-s\right)  \ ,\\
\Psi_{\ast}\left(  x,s\right)   &  =\frac{2-s}{s}~\Psi_{0}\left(  \frac
{sx}{2-s},2-s\right)  \ ,\\
\Psi_{g}\left(  x,s\right)   &  =\frac{1-s}{2-s}+\Psi_{\ast}\left(
x+\frac{s-1}{2-s},2-s\right) \\
\Psi_{\ast}\left(  x,s\right)   &  =\frac{1-s}{s}+\Psi_{g}\left(  x+\frac
{s-1}{s},2-s\right) \\
\Psi_{g}\left(  x,s\right)   &  =\frac{1-s}{2-s}+\frac{s}{2-s}\Psi_{0}\left(
\frac{s-1+\left(  2-s\right)  x}{s},s\right)  \ ,\\
\Psi_{0}\left(  x,s\right)   &  =\frac{s-1}{s}+\frac{2-s}{s}\Psi_{g}\left(
\frac{1-s+sx}{2-s},s\right)  \ ,
\end{align}
For the potential itself, functional conjugation leads to%
\end{subequations}
\begin{align}
V\left(  x,s\right)   &  =\left(  \frac{2-s}{s}\right)  ^{2}V\left(  \frac
{s}{2-s}\left(  x-\left(  1-\frac{1}{s}\right)  \right)  ,2-s\right)  \ ,\\
U\left(  x,s\right)   &  =\left(  \frac{\left(  2-s\right)  \ln\left(
2-s\right)  }{s\ln s}\right)  ^{2}U\left(  \frac{s}{2-s}\left(  x-\left(
1-\frac{1}{s}\right)  \right)  ,2-s\right)  \ .
\end{align}

Perhaps it is worthwhile to work out the various functions explicitly for the
cases which can be solved in closed-form, with $\Psi_{0}$\ and $\Phi_{0}%
=\Psi_{0}^{-1}$\ as given in (\ref{ExactCases}). \ Consider the first and the
last of these. \ The corresponding dual and conjugated functions are given by:%
\begin{subequations}
\begin{align}
\Psi_{\ast}\left(  x,4\right)   &  =-\frac{1}{2}\Psi_{0}\left(  -2x,-2\right)
=-\frac{\sqrt{3}}{12}\left(  2\pi-3\arccos\left(  2x+\frac{1}{2}\right)
\right)  \ ,\\
\Psi_{\ast}\left(  x,-2\right)   &  =-2\Psi_{0}\left(  -\frac{1}{2}x,4\right)
=-2\left(  \arcsin\sqrt{-\frac{1}{2}x}\right)  ^{2}\ ,\\
\Psi_{g}\left(  x,4\right)   &  =\frac{3}{2}+\Psi_{\ast}\left(  x-\frac{3}%
{2},-2\right)  =\frac{3}{2}-2\left(  \arcsin\sqrt{\frac{1}{2}\left(  \frac
{3}{2}-x\right)  }\right)  ^{2}\ ,\\
\Psi_{g}\left(  x,-2\right)   &  =\frac{3}{4}+\Psi_{\ast}\left(  x-\frac{3}%
{4},4\right)  =\frac{3}{4}-\frac{\sqrt{3}}{12}\left(  2\pi-3\arccos\left(
2x-1\right)  \right)  \ .
\end{align}
We plot these, say for $x$ such that the arguments of the inverse
trigonometric functions are real, and the functions are principal-valued.
\end{subequations}
\begin{center}%
{\parbox[b]{3.4911in}{\begin{center}
\includegraphics[
height=2.3263in,
width=3.4911in
]%
{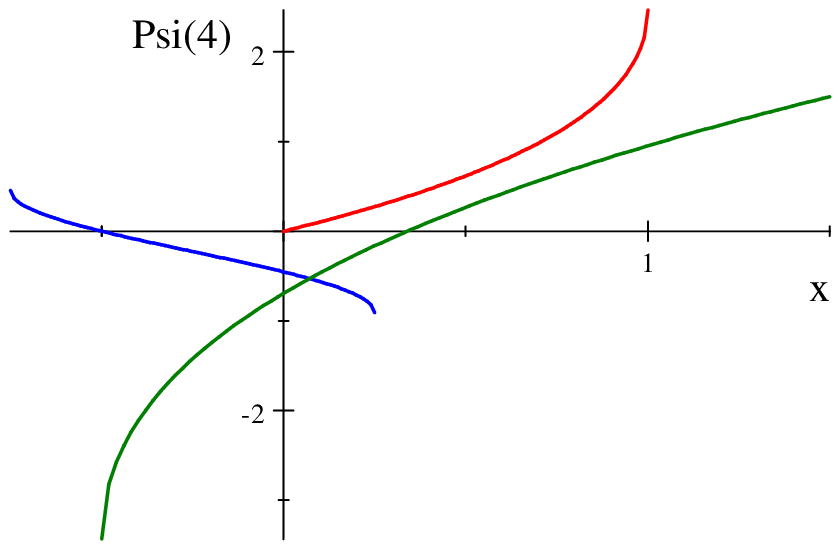}%
\\
$\Psi_{0}\left(  x,4\right)  $ red, $\Psi_{\ast}\left(  x,4\right)  $ blue,
and $\Psi_{g}\left(  x,4\right)  $ green.
\end{center}}}%
{\parbox[b]{3.506in}{\begin{center}
\includegraphics[
height=2.3263in,
width=3.506in
]%
{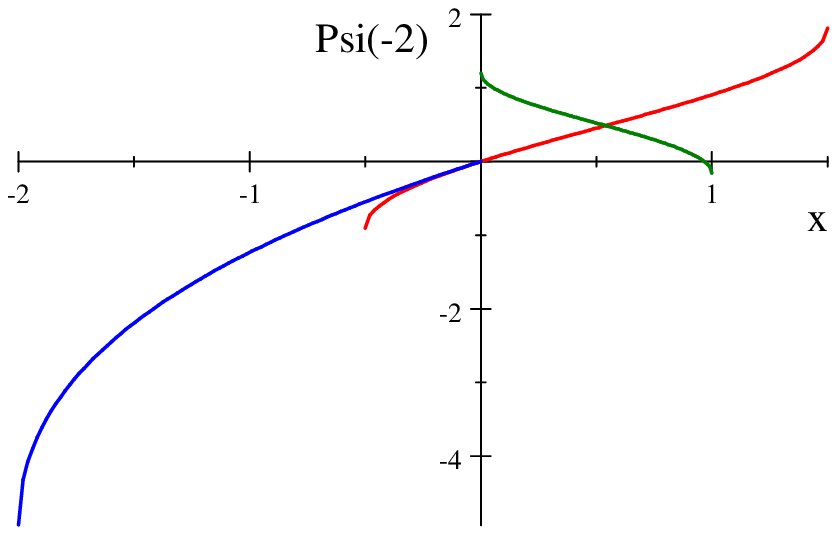}%
\\
$\Psi_{0}\left(  x,-2\right)  $ red, $\Psi_{\ast}\left(  x,-2\right)  $ blue,
and $\Psi_{g}\left(  x,-2\right)  $ green.
\end{center}}}%

\end{center}

\subsection{Alternate series solution theorem}

Consider the Poincar\'{e} functional equation for the Schr\"{o}der function
inverse, rendered to facilitate expansion about the nontrivial fixed point
\textquotedblleft$\ast$\textquotedblright\ corresponding to $x_{\ast}=1-1/s$,
\
\begin{equation}
\Phi_{\ast}\left(  \left(  2-s\right)  z,s\right)  =\left(  2-s\right)
\Phi_{\ast}\left(  z,s\right)  -s\Phi_{\ast}^{2}\left(  z,s\right)  \text{
\ \ with \ \ }\Phi_{\ast}\left(  z,s\right)  =z+O\left(  z^{2}\right)  \ ,
\label{PoincareStar}%
\end{equation}
where%
\begin{equation}
x=1-\frac{1}{s}+z\ ,\ \ \ z=x+\frac{1}{s}-1\ .
\end{equation}
Let
\begin{equation}
\Phi_{\ast}\left(  z,s\right)  =\frac{2-s}{s}~\phi\left(  z,s\right)  \ ,
\end{equation}
to change the functional equation to%
\begin{equation}
\frac{2-s}{s}~\phi\left(  \left(  2-s\right)  z,s\right)  =\frac{\left(
2-s\right)  ^{2}}{s}~\phi\left(  z,\lambda\right)  \left(  1-\phi\left(
z,\lambda\right)  \right)  \ .
\end{equation}
That is to say,%
\begin{equation}
\phi\left(  \left(  2-s\right)  z,s\right)  =\left(  2-s\right)  \phi\left(
z,s\right)  \left(  1-\phi\left(  z,s\right)  \right)  \text{ \ \ with
\ \ }\phi\left(  z,s\right)  =\frac{sz}{2-s}+O\left(  z^{2}\right)  \ .
\end{equation}
Comparing this last relation to the original Poincar\'{e} functional equation
tailored to yield the expansion about the trivial fixed point
\textquotedblleft$0$\textquotedblright,%
\begin{equation}
\Phi_{0}\left(  sx,s\right)  =s\Phi_{0}\left(  x,s\right)  \left(  1-\Phi
_{0}\left(  x,s\right)  \right)  \text{ \ \ with \ \ }\Phi_{0}\left(
x,s\right)  =x+O\left(  x^{2}\right)  \ , \label{PoincareZero}%
\end{equation}
it follows that
\begin{equation}
\phi\left(  z,s\right)  =\Phi_{0}\left(  \frac{sz}{2-s},2-s\right)  \ .
\end{equation}
So then%
\begin{equation}
\Phi_{\ast}\left(  z,s\right)  =\frac{2-s}{s}~\Phi_{0}\left(  \frac{sz}%
{2-s},2-s\right)  \ ,\ \ \ \Phi_{0}\left(  z,s\right)  =\frac{2-s}{s}%
~\Phi_{\ast}\left(  \frac{sz}{2-s},2-s\right)  \ , \label{PhiStarPhiZero}%
\end{equation}
and therefore the Schr\"{o}der function, i.e. the inverse of the inverse
function $\Psi=\Phi^{-1}$, is%
\begin{equation}
\Psi_{\ast}\left(  z,s\right)  =\frac{2-s}{s}~\Psi_{0}\left(  \frac{sz}%
{2-s},2-s\right)  \ ,\ \ \ \Psi_{0}\left(  z,s\right)  =\frac{2-s}{s}%
~\Psi_{\ast}\left(  \frac{sz}{2-s},2-s\right)  \ . \label{PsiStarPsiZero}%
\end{equation}
Hence $\Phi_{\ast}\left(  \Psi_{\ast}\left(  z,s\right)  ,s\right)
=z=\Psi_{\ast}\left(  \Phi_{\ast}\left(  z,s\right)  ,s\right)  $. \ This
proves the relationships between $\Phi_{\ast},\Psi_{\ast}$ and $\Phi_{0}%
,\Psi_{0}$. \ $\blacksquare$

\section{Appendix B: \ Constructing the branches of Schr\"{o}der's $\Psi$
function}

We wish to imitate for $\Psi\left(  x,s\right)  $ what we did to find the
branches of the potential in Section III of the text. \ We start with the
functional equation, (\ref{SFE}). \ Next we write $y=sx\left(  1-x\right)  $,
with solutions $x_{\pm}=\frac{1}{2}\left(  1\pm\sqrt{1-4y/s}\right)  $. \ Now,
we rename $y\rightarrow x$, so the functional relation becomes%
\begin{equation}
\Psi_{\pm}\left(  x,s\right)  =s\Psi\left(  \frac{1}{2}\left(  1\pm
\sqrt{1-4x/s}\right)  ,s\right)  \ .
\end{equation}
One of these ($\Psi_{-}$) reproduces the original auxiliary function expanded
about $x=0$, only it does so more accurately for $x\rightarrow s/4$ as may be
seen by numerical evaluation, while the other ($\Psi_{+}$) gives the auxiliary
on another sheet of the function's Riemann surface. \ This is useful for
determining the first switchback potential. \ The process may be repeated to
get the other branches of the auxiliary as a sequence of functions, $\Psi_{n}$.

Here we are especially interested in $s=10/3$, as we wish to confirm the order
in which the switchback potentials are encountered by the evolving particle
for this case. \ We find the following numerical results.%
\begin{center}
\includegraphics[
trim=0.000000in 0.000000in 0.000000in -0.005616in,
height=6.5039in,
width=4.4998in
]%
{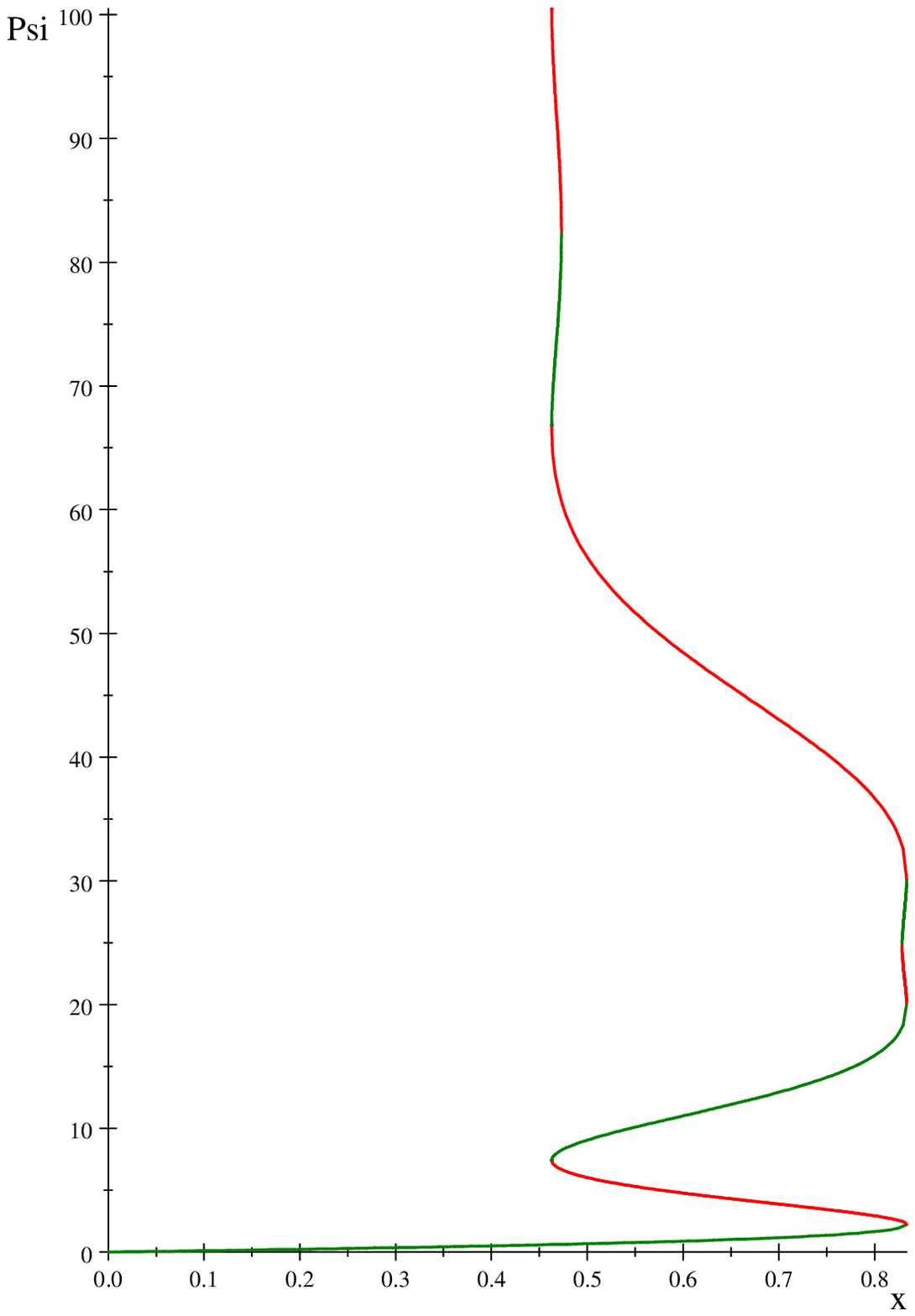}%
\\
Eight branches of $\Psi\left(  x,10/3\right)  $.
\end{center}
It follows that the order of the potentials is as given in the text:
\ $V_{0}\rightarrow V_{1}\rightarrow V_{2}\rightarrow W_{1}\rightarrow
W_{2}\rightarrow V_{3}\rightarrow X_{1}\rightarrow X_{2}\rightarrow\cdots$.

As a practical matter, it is somewhat easier to compute the inverse function
$\Phi\equiv$ $\Psi^{-1}$ by combining series solution methods with functional
extensions. \ The functional equation for the inverse of Schr\"{o}der's
function is%
\begin{equation}
\Phi\left(  sx,s\right)  =s\Phi\left(  x,s\right)  \left(  1-\Phi\left(
x,s\right)  \right)  \ . \label{Poincare}%
\end{equation}
For $s<1$ this quadratic equation may be solved for $\Phi\left(  x\right)  $
in terms of $\Phi\left(  sx\right)  $,
\begin{equation}
\Phi_{\pm}\left(  x,s\right)  =\frac{1}{2}\left(  1\pm\sqrt{1-\frac{4}{s}%
\Phi\left(  sx,s\right)  }\right)  \ , \label{Phi+-}%
\end{equation}
and, upon iteration, all values of $\Phi$ may be obtained from those values
given accurately by the series about $x=0$. \ For $s>1$, on the other hand, we
first rescale $x$ in (\ref{Poincare}) to write%
\begin{equation}
\Phi\left(  x,s\right)  =s\Phi\left(  \frac{x}{s},s\right)  \left(
1-\Phi\left(  \frac{x}{s},s\right)  \right)  \ , \label{PhiExtension}%
\end{equation}
and from this equation, upon iteration, all values of $\Phi$ may be obtained
directly from those values given accurately by the series about $x=0$. \ 

As an example, consider again the case $s=10/3$. \ We find:%
\begin{center}
\includegraphics[
height=4.5155in,
width=6.5471in
]%
{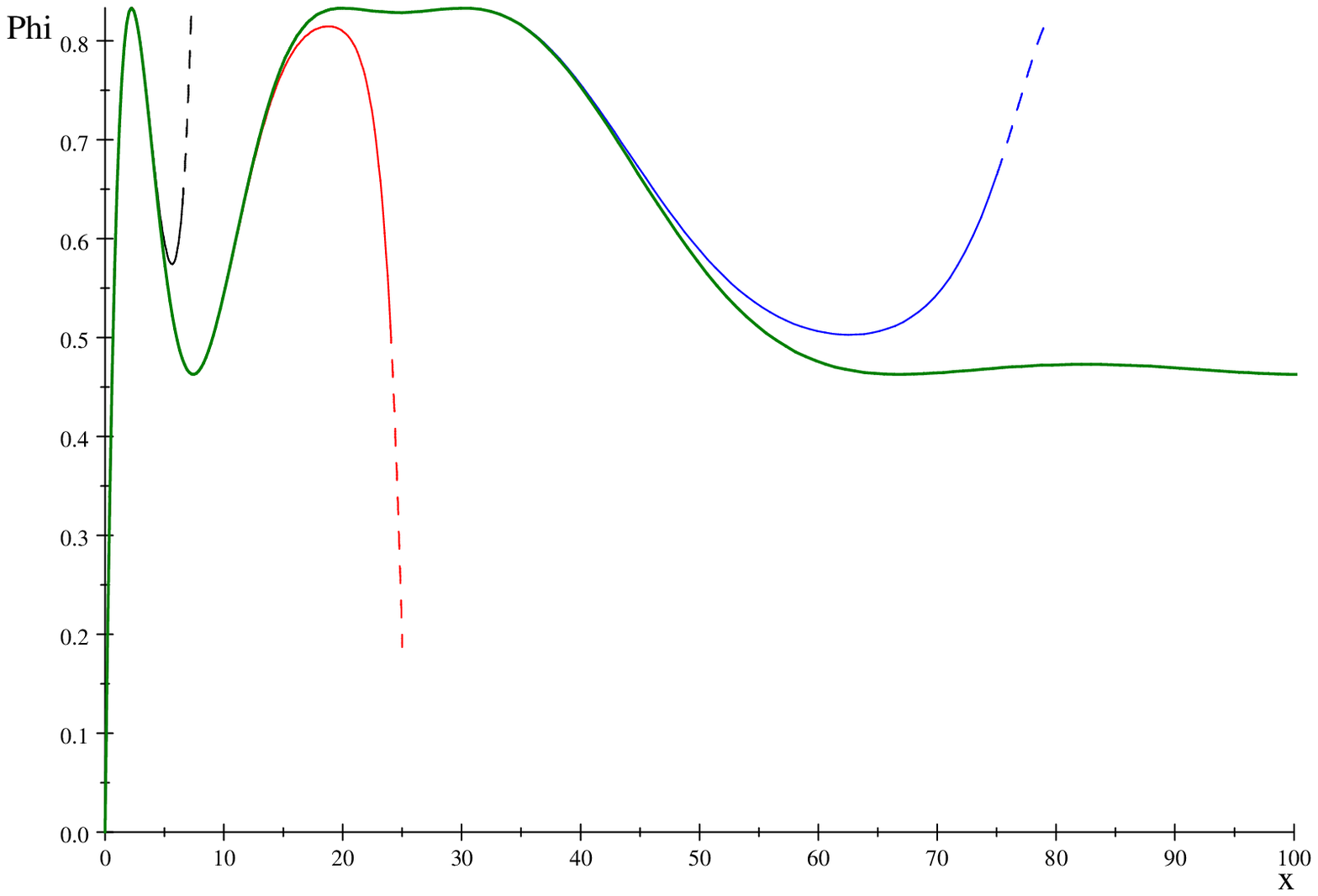}%
\\
Seventh order series approximation to $\Phi\left(  x,10/3\right)  $, in black,
compared to results from using this series in the 1st (red), 2nd (blue), and
3rd (green) iterates of the functional extension (\ref{PhiExtension}).
\end{center}
Flipping this graph about the SW-NE diagonal, we obtain a graph for the
multi-valued $\Psi=\Phi^{-1}$. \ From the third iterate of (\ref{PhiExtension}%
) (as given in the graph by the green curve) the resulting curve for
$\Psi\left(  x,10/3\right)  $ agrees with the previous graph obtained by
solving Schr\"{o}der's equation through use of the series about $x=0$,
combined with functional methods.

\section{Appendix C: \ The $s=1$ series for the potential}

For this case the appropriate limit of (\ref{USeries}) gives%
\begin{align}
V\left(  x,1\right)   &  \equiv-\lim\limits_{s\rightarrow1}\left(  \left(
\ln^{2}s\right)  U\left(  x,s\right)  \right) \nonumber\\
&  =-x^{4}-2x^{5}-4x^{6}-\frac{25}{3}x^{7}-\frac{215}{12}x^{8}-\frac{589}%
{15}x^{9}-\frac{7813}{90}\allowbreak x^{10}-\frac{60\,481}{315}x^{11}%
-\frac{11\,821}{28}x^{12}+O\left(  x^{13}\right)  \ . \label{V1}%
\end{align}
More systematically, let%
\begin{equation}
V\left(  x,1\right)  =-x^{4}\left(  1+\sum_{n=1}^{\infty}c_{n}~x^{n}\right)
\ ,\ \ \ c_{1}=2\ ,\ \ \ c_{2}=4\ ,\ \ \ c_{3}=\frac{25}{3}\ ,\ \cdots\ ,
\label{V(s=1)Series}%
\end{equation}
and solve by iteration the functional equation that should be obeyed by
$V\left(  x,1\right)  $, namely,%
\begin{equation}
V\left(  x\left(  1-x\right)  ,1\right)  =\left(  1-2x\right)  ^{2}V\left(
x,1\right)  \ .
\end{equation}
If the formal series (\ref{V(s=1)Series}) is constructed to $O\left(
x^{25}\right)  $ or so, it appears to have a radius of convergence of
$R\simeq1/2$. \ But\ remarkably, this is seen to be illusory to higher orders.
\ Different behavior sets in around $O\left(  x^{30}\right)  $, where the
successive $\left\vert c_{n}\right\vert ^{1/n}$ used in the $\lim\sup$
determination of $R$, (\ref{Radii}), begin to grow linearly for $n>30$, as
shown here (purple curve, with wiggles).%
\begin{center}
\includegraphics[
height=3.9693in,
width=6.3678in
]%
{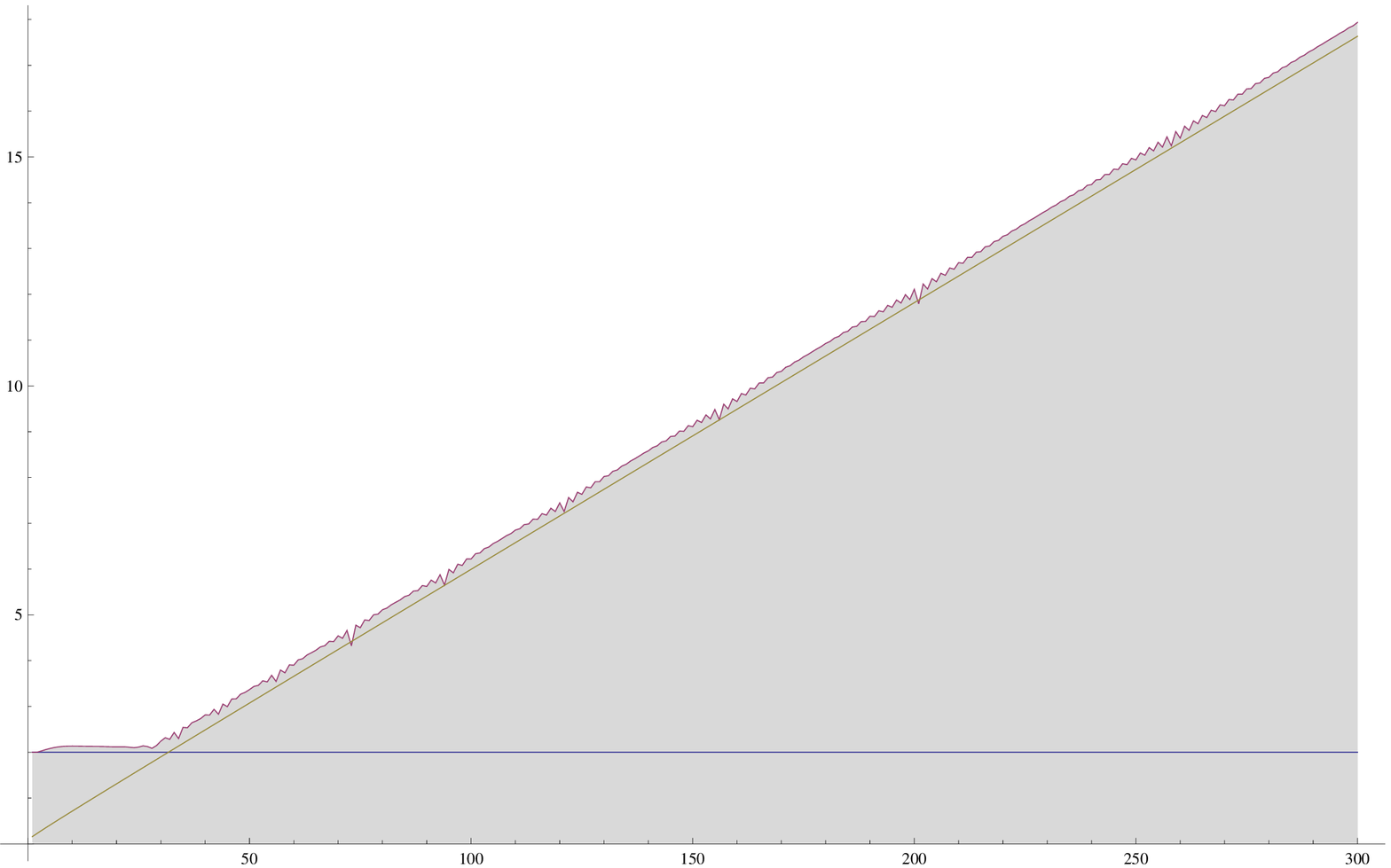}%
\\
$\left\vert c_{n}\right\vert ^{1/n}$ for $s=1$ and $1\leq n\leq300$. \ The
blue horizontal is\ $2$, while the sienna line is $\left(  2^{-n/2}%
e^{-3n/2}n!\right)  ^{1/n}$.
\end{center}
That is to say, $\left\vert c_{n}\right\vert \sim L^{n}\exp\left(  n\ln
n\right)  $ for large $n$, with $L\simeq\frac{16}{270}\simeq\frac{1}{\sqrt
{2}e^{5/2}}=\allowbreak5.\,\allowbreak804\,29\times10^{-2}$. \ This
immediately brings to mind Stirling's formula, $n!\sim\sqrt{2\pi n}\left(
\frac{n}{e}\right)  ^{n}$, and in fact, a comparison of $c_{n}$ with
$f_{n}\equiv2^{-n/2}e^{-3n/2}n!$ is striking, as shown in the Figure. \ (The
most important feature in the Figure is agreement between the averaged slopes.
\ A better overall fit to the $c_{n}$, on average, is achieved by shifting the
sienna line slightly to the left, for example by including an additional
$\sqrt{n}$ factor in $f_{n}$.) \ 

All this is compelling evidence that (\ref{V(s=1)Series}) itself is \emph{not}
convergent, but rather an asymptotic series. \ Indeed, the $f_{n}$
\textquotedblleft fit\textquotedblright\ to $c_{n}$ can be obtained from an
asymptotic approximation of a simple integral:%
\begin{align}
I\left(  x\right)   &  \equiv\int_{0}^{\infty}e^{-y}\frac{1}{1-\frac{xy}%
{\sqrt{2e^{3}}}}dy=-\frac{\sqrt{2}}{x}~\exp\left(  \frac{3x-2\sqrt{2}%
e^{\frac{3}{2}}}{2x}\right)  \operatorname{Ei}\left(  1,-\frac{1}{x}\sqrt
{2}e^{\frac{3}{2}}\right) \nonumber\\
&  =\int_{0}^{\infty}e^{-y}\sum_{n=0}^{\infty}x^{n}2^{-n/2}e^{-3n/2}%
y^{n}dy\sim\sum_{n=0}^{\infty}f_{n}~x^{n}\ .
\end{align}
Taking the Cauchy principal value for $I\left(  x\right)  $ gives finite
numerical results for all $x$. \ These results may be used to compute
corrections to the polynomial approximation for $V\left(  x,1\right)  $ as
constructed from truncating the series (\ref{V(s=1)Series}). \ We leave
further analysis of this interesting but peculiar case to the reader.

\end{document}